\documentclass[preprint]{aastex63}
\usepackage{amsmath}
\begin{document}

\title{Extreme Ultraviolet Late Phase of Solar Flares}

\author{Jun Chen}
\affiliation{CAS Key Laboratory of Geospace Environment, Department of Geophysics and Planetary Sciences, University of Science and Technology of China, Hefei 230026, China}
\affiliation{Institute of Physics and Astronomy, University of Potsdam, 14476 Potsdam, Germany}

\author[0000-0003-4618-4979]{Rui Liu}
\affiliation{CAS Key Laboratory of Geospace Environment, Department of Geophysics and Planetary Sciences, University of Science and Technology of China, Hefei 230026, China}
\affiliation{CAS Center for Excellence in Comparative Planetology, Hefei 230026, China}

\author{Kai Liu}
\affiliation{CAS Key Laboratory of Geospace Environment, Department of Geophysics and Planetary Sciences, University of Science and Technology of China, Hefei 230026, China}
\affiliation{Mengcheng National Geophysical Observatory, School of Earth and Space Sciences, University of Science and Technology of China, Hefei 230026, China}

\author[0000-0001-5313-1125]{Arun Kumar Awasthi}
\affiliation{CAS Key Laboratory of Geospace Environment, Department of Geophysics and Planetary Sciences, University of Science and Technology of China, Hefei 230026, China}
\affiliation{CAS Center for Excellence in Comparative Planetology, Hefei 230026, China}

\author[0000-0001-6855-5799]{Peijin Zhang}
\affiliation{CAS Key Laboratory of Geospace Environment, Department of Geophysics and Planetary Sciences, University of Science and Technology of China, Hefei 230026, China}

\author[0000-0002-8887-3919]{Yuming Wang}
\affiliation{CAS Key Laboratory of Geospace Environment, Department of Geophysics and Planetary Sciences, University of Science and Technology of China, Hefei 230026, China}
\affiliation{CAS Center for Excellence in Comparative Planetology, Hefei 230026, China}

\author[0000-0002-5740-8803]{Bernhard Kliem}
\affiliation{Institute of Physics and Astronomy, University of Potsdam, 14476 Potsdam,
Germany}

\correspondingauthor{Rui Liu}
\email{rliu@ustc.edu.cn}
\begin{abstract}
	A second peak in extreme ultraviolet sometimes appears during the gradual phase of solar flares, which is known as EUV late phase (ELP). Stereotypically ELP is associated with two separated sets of flaring loops with distinct sizes, and it has been debated whether ELP is caused by additional heating or extended plasma cooling in the longer loop system. Here we carry out a survey of 55 M-and-above GOES-class flares with ELP during 2010--2014. Based on the flare-ribbon morphology, these flares are categorized as circular-ribbon (19 events), two-ribbon (23 events), and complex-ribbon (13 events) flares. Among them, 22 events (40\%) are associated with coronal mass ejections, while the rest are confined. An extreme ELP, with the late-phase peak exceeding the main-phase peak, is found in 48\% of two-ribbon flares, 37\% of circular-ribbon flares, and 31\% of complex-ribbon flares, suggesting that additional heating is more likely present during ELP in two-ribbon than in circular-ribbon flares. Overall, cooling may be the dominant factor causing the delay of the ELP peak relative to the main-phase peak, because the loop system responsible for the ELP emission is generally larger than, and well separated from, that responsible for the main-phase emission. All but one of the circular-ribbon flares can be well explained by a composite ``dome-plate'' quasi-separatrix layer (QSL). Only half of these show a magnetic null point, with its fan and spine embedded in the dome and plate, respectively. The dome-plate QSL, therefore, is a general and robust structure characterizing circular-ribbon flares. 
\end{abstract}


\section{Introduction}

Solar flares are localized, transient brightenings on the Sun \citep{Fletcher2011}. The flare emission increases across the entire electromagnetic spectrum; but often the GOES soft X-ray (SXR) flux is used as the flare proxy: it increases during the so-called ``impulsive phase'' on time-scales of seconds to minutes, and then gradually decays on time-scales of minutes to tens of minutes up to several hours, hence termed ``gradual phase''. Generally extreme-ultraviolet (EUV) emissions peak sequentially in an order of decreasing temperatures shortly after the SXR peak, which is often termed ``main phase''. Recently, using the full-disk integrated EUV irradiance observed by the EUV Variability Experiment \citep[EVE;][]{Woods2012} onboard the \textsl{Solar Dynamics Observatory} \citep[\textsl{SDO};][]{Pesnell2012}, \citet{Woods2011} discovered that some flares exhibit an additional peak in EUV ``warm'' coronal lines (e.g.  \ion{Fe}{16} 335~{\AA}, $\sim\,$2.5 MK) several tens of minutes to hours after the SXR peak, which is termed ``EUV late phase'' (ELP). Fluctuations in EUV irradiance can drive immediate changes in the Earth's upper atmosphere, which has significant space-weather consequences, such as compromised satellite lifespan, radio communication, and satellite navigation. The effects of EUV irradiance have also been noted on Mars \citep[e.g.,][]{Withers2009} and on the Moon \citep[e.g.,][]{Sternovsky2008}. Hence the cause of ELP, which is still poorly understood, has raised great interest.

\citet{Woods2011} found that only a small fraction (13\%) of 191 C2-and-above flares exhibit ELP. About half of these ELP events occur in a cluster of two active regions, implying a multipolar configuration. In case studies using spatially resolved imaging observations, it has been found that the ELP emission comes from a set of loops higher and longer than the flaring loops observed during the main phase \citep[e.g.,][]{Woods2011,Hock2012,KaiLiu2013,DaiYu2013,Sun2013,Masson2017,DaiYu2018}. Almost exclusively, these ELP flares have a circular-shaped ribbon associated with the main-phase flare arcade and a remote ribbon associated with the ELP arcade. Often the fan-spine topology of a magnetic null point links the main-phase and ELP flare arcades: the quasi-circular flare ribbon corresponds to the footprint of the fan and the ELP loops are associated with the spine \citep{Sun2013,LiY2014,DaiYu2018}. However, since the remote ribbon is extended rather than pointwise, the surrounding quasi-separatix layers \citep[QSLs;][]{Priest&Demoulin1995} must also be involved in the reconnection process \citep{Reid2012,Masson2017,DaiYu2018}.  

The so-called ``extreme ELP'' events have a higher ELP peak than the main-phase peak at EVE 335~{\AA} \citep[e.g.,][]{KaiLiu2015,DaiYu2018,Zhou2019}, which argues for additional heating during the gradual phase. An alternative scenario is that both the main-phase loops and ELP loops are heated during the main phase, but the delayed ELP results from an extended plasma cooling process \citep{KaiLiu2013,Masson2017}. This is because the conductive cooling time increases with longer loop length, while the radiative cooling time increases with lower density in the longer and higher ELP loop. Combining the two factors leads to significantly different cooling rates between loops of different lengths. It is possible that both mechanisms are at work, if ELP loops are associated with the hot spine of a magnetic null \citep{Sun2013,LiY2014} or when magnetic reconnection continually occurs during the gradual phase \citep{Zhou2019}. In particular, it is suggested that a flux rope energized but later trapped in confined eruptions may provide persistent heating during the gradual phase \citep{KaiLiu2015,Wang2016}, presumably through magnetic reconnection at the rope's QSL boundary \citep{Liu2016apj,Wang2017} or Joule heating induced by internal kink instability \citep{Galsgaard+Nordlund1997}. Assuming either long-lasting plasma cooling or additional heating, some authors \citep[e.g.,][]{Sun2013,LiY2014,Dai+Ding2018,DaiYu2018,Zhou2019} are able to reproduce ELP-like light curves with the Enthalpy-Based Thermal Evolution of Loops (EBTEL) model \citep{Cargill2012}. Hence, it is inconclusive which mechanism is dominant, although it was suggested that additional heating may manifest itself as peculiar features in light curves \citep[e.g.,][]{LiY2014,Dai+Ding2018}.

Therefore, except for a few carefully studied events, questions remains open as to where the ELP emission originates, whether ELP is always associated with magnetic nulls, or how relatively important heating and cooling are in the production of ELP. To shed light on these questions, we carried out a survey of major ELP flares (M- and X-class) recorded by EVE and investigated the characteristics of these flares with emphasis on the flare morphology as manifested by flare ribbons and loops and on the relevant magnetic topology. The rest of the paper are organized as follows. The instruments and methods used in this study are briefly introduced in \S\ref{Method}, the results are presented in \S\ref{sec:results}, and concluding remarks are given in \S\ref{sec:conclusion}.

\section{Instruments \& Methods}\label{Method}

There are 473 M- and X- class flares during the period from May 2010 to May 2014. Among them we selected 55 flares (Appendix~\ref{appendix:flarelist}) in this study according to two criteria as follows. 1) A selected flare must possess a significant ELP in the EVE 335~\AA\ irradiance profile. 2) The flaring region must be located within $45^{\circ}$ from the disk center, so that reliable measurements of photospheric magnetic field are available. We identified flares with ELP using \ion{Fe}{16} 335~{\AA} of EVE level-2 line data with a temporal cadence of 10 s and an accuracy of 20\% \citep{Woods2012}. We then determined the flare source regions using EUV/UV images taken by the Atmospheric Imaging Assembly \citep[AIA;][]{Lemen2012} onboard \textsl{SDO}. AIA provides full-disk images with a pixel scale of $0.6''$ and a cadence of 12~s for EUV and of 24~s for UV passbands. Integrating over the flare region, we made EUV lightcurves with different AIA passbands. By comparing the AIA 335~{\AA} lightcurve with the EVE 335~{\AA} irradiance, we confirmed that each selected flare is the source of the ELP, and we further made sure that unlike the main phase, the late-phase 335~{\AA} peak has no counterpart in GOES soft X-rays or AIA 131~{\AA} (primarily \ion{Fe}{21} for flare, peak response temperature $\log T = 7.05$), i.e., it is unlikely a second flare in the same region. However, we did not exclude those events in which a small flare (below M-class) occurs in the same region following the major flare, but well before the late-phase peak. In this case, we made an assumption that the major flare makes a major contribution to the ELP when considering the time delay of the ELP peak relative to the main-phase peak (see \S\ref{subsec:stat}). 

Compared with the original criteria adopted by \citet{Woods2011} to identify ELP flares, we have relaxed two restrictions, i.e., we did not require an ELP to be associated with an eruptive flare, nor demand in advance the presence of a second set of EUV loops that are longer and higher and appear much later than the main-phase flare loops, because these restrictions are irrelevant to the space-weather effects of EUV irradiance. Instead, we will examine below (\S\ref{subsec:stat}) whether the collected events tend to be eruptive and whether their ELP loops tend to be separated from main-phase loops.

To understand the magnetic configuration of the flare source regions, we performed potential-field extrapolation with the Fourier transform method \citep{Alissandrakis1981}. Potential field maintains basic structural skeletons \citep{Titov2007}, whose robustness has been demonstrated by earlier studies employing various coronal field models \citep[e.g.,][]{Liu2014,Liu2016apj,Liu2016sr}. The extrapolation is based on the Space-Weather HMI Active Region Patches (SHARP; \texttt{hmi.sharp\_cea}) data series obtained by the Helioseismic and Magnetic Imager \citep[HMI;][]{Hoeksema2014} onboard \textsl{SDO}. These vector magnetograms were disambiguated and deprojected to the heliographic coordinates with a Lambert (cylindrical equal area, CEA) projection method, resulting in a pixel scale of 0.36 Mm \citep{Bobra2014}. Before extrapolation, we ``pre-processed'' a pre-flare vector magnetogram to push it towards being force-free \citep{Wiegelmann2006}. With the extrapolated potential field, we calculated the squashing factor $Q$ \citep{Titov2002} to quantify magnetic connectivity \citep[for details see][]{Liu2016apj}, and also searched for null points with an iterative Newton-Raphson method \citep{Haynes&Parnell2007}. 

\section{Results}\label{sec:results}

Based on the flare-ribbon morphology observed in the AIA 1600~{\AA} passband, we categorized the ELP flares selected in this study as circular-ribbon, two-ribbon, and complex-ribbon flares. With elongated ribbons on both sides of the magnetic polarity inversion line, two-ribbon flares are well explained by the standard flare model \citep{standard1,standard2,standard3,standard4}. 
Circular-ribbon flares are often associated with the fan-spine topology of a three-dimensional (3D) coronal null point \citep{Wang+Liu2012}. 
The rest are termed complex-ribbon flares because the ribbon morphology is complicated and often has no distinct features. Below we analyze an exemplary circular-ribbon (\S\ref{subsec:cr}) and an exemplary two-ribbon flare (\S\ref{subsec:2r}), and then present the statistical results for the sample of 55 ELP flares (\S\ref{subsec:stat}).

\subsection{Characteristics of a circular-ribbon ELP flare}\label{subsec:cr}

\begin{figure}[htbp]
	\centering
	\plotone{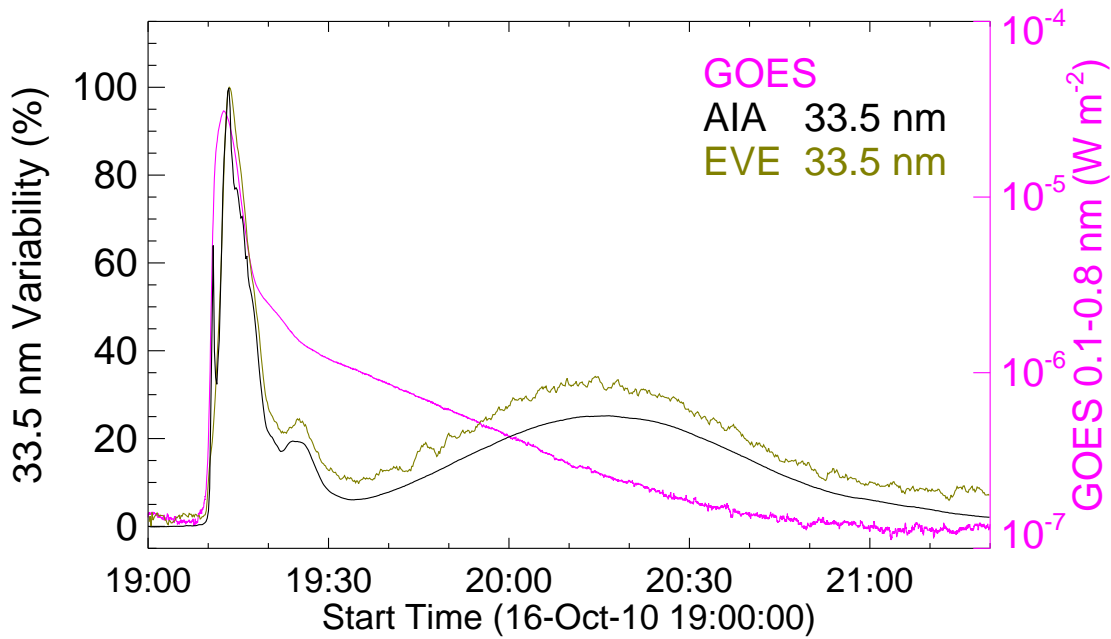}
	\caption{Light curves of the \textsl{GOES} M2.9-class flare on 2010 October 16. \textsl{GOES} 1-8~\AA\ flux (magenta; scaled by the right $y$-axis), EVE irradiance at 335~{\AA} (\ion{Fe}{16}; tan), and AIA 335~{\AA} data number (black) integrated over the ELP region identified in Figure~\ref{fig:2010-10-16diff335}a. The 335~{\AA} light curves are scaled by the left $y$-axis and normalized by setting the value at 19:00:00~UT to be 0 and the maximum value to be 1.	}
	\label{fig:2010-10-16profile}
\end{figure}

\begin{figure}[htbp]
	\centering
	\plotone{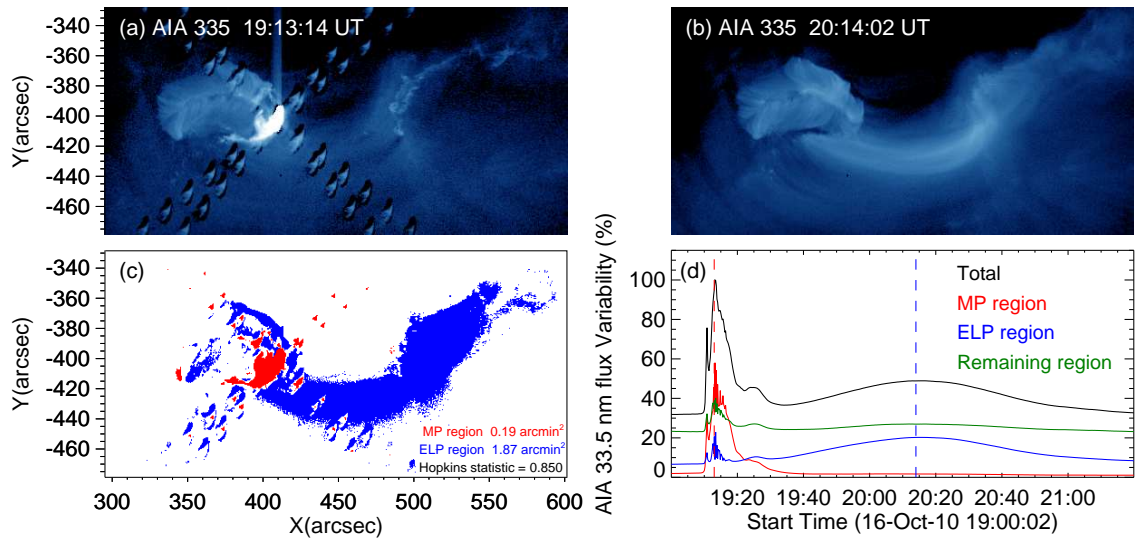}
	\caption{Flaring regions of the ELP flare on 2010 October 16. (a) AIA 335~{\AA} image taken at the main-phase peak.
		(b) AIA 335~{\AA} image taken at the ELP peak, with the same field of view as (a). (c) Main phase (ELP) region indicated by blue (red) area.
		(d) Temporal variation of  AIA 335~{\AA} data number integrated over the regions in (c); the vertical dashed red (blue) line marks the times of images in Panels (a) and (b).} 	\label{fig:2010-10-16diff335}
\end{figure}

Here we present an exemplary circular-ribbon flare, an M2.9 class flare occurring in NOAA active region (AR) 11112 on 2010 October 16. The SXR light curve indicates that the flare begins at $\sim$19:07 UT and peaks at $\sim$19:12 UT (Figure~\ref{fig:2010-10-16profile}). During the gradual phase, an additional enhancement in \ion{Fe}{16} 335 \AA\ from $\sim\,$19:34 UT to $\sim\,$21:00 UT is recognized as an ELP in \citet{KaiLiu2013}. During this event, there is no other major activity on the disk. The light curve of the total flux from the field of view (FOV) in Figure~\ref{fig:2010-10-16diff335}(a-c) at AIA 335 \AA\ has a similar profile as the EVE 335 \AA\ irradiance, confirming that the flare possesses an ELP.

The emission area of the main phase and the ELP is estimated as follows. The AIA 335~\AA\ image at the main-phase peak (Figure~\ref{fig:2010-10-16diff335}a) is subtracted by the image at the late-phase peak (Figure~\ref{fig:2010-10-16diff335}b). In the difference image, the main-phase region is identified as those pixels whose values are above the average of all pixels with positive values, and the late-phase region as those below the average of all pixels with negative values (Figure~\ref{fig:2010-10-16diff335}c). Unfortunately, the main-phase image is contaminated by the CCD bleeding and diffraction patterns, even after the point spread function deconvolution using the SolarSoftware procedure \texttt{aia\_deconvolve\_richardsonlucy.pro} (Figure~\ref{fig:2010-10-16diff335}a). Despite the contamination, the light curve of the identified main-phase region has a major peak similar to the light curve from the total FOV during the main phase, whereas being flat during the ELP; the reverse is true for the light curve of the identified late-phase region (Figure~\ref{fig:2010-10-16diff335}d). Thus, we have basically separated the two regions with this simple approach. To quantify the separation, we calculate the Hopkins statistic (Appendix~\ref{appendix:hopkins}), which gives a value as high as 0.850 to be consistent with a significant separation.

\begin{figure}[htbp]
\centering
 \plotone{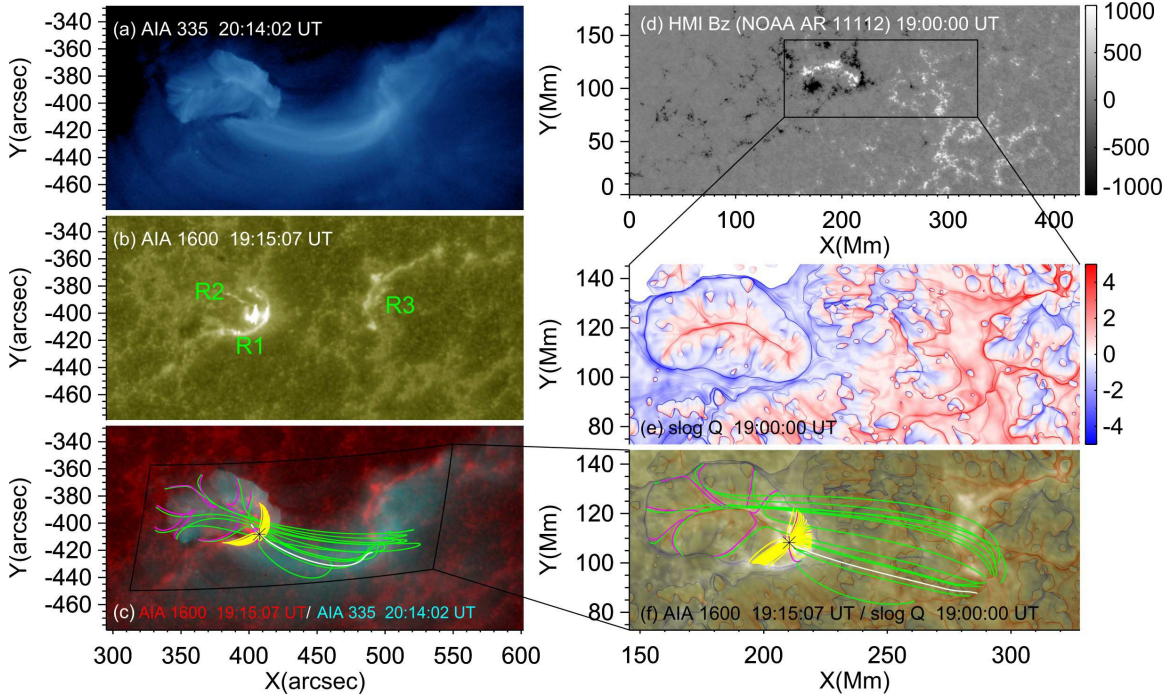}
\caption{Magnetic configuration of the ELP flare on 2010 October 16. (a) AIA 335~{\AA} image taken at the ELP peak. (b) AIA 1600~{\AA} image taken close to the main-phase peak. (c) Composite image showing AIA 335~{\AA} in red at 20:14:05 UT and 1600~{\AA} in cyan at 19:15:02 UT. (d) $B_z$ component of the photospheric magnetic field of NOAA AR 1112 at 19:00~UT in the CEA coordinates; the rectangle indicates the FOV of Panels (e) and (f) and corresponds to the curved parallelogram in Panel (d). (e) Map of signed $\log Q$, $\mathrm{slog}\,Q\equiv\mathrm{sign}(Bz)\,\log Q$. (f) A blend of $\log Q$-map and AIA 1600~{\AA} at 19:15:02~UT. Spine and fan of a null point (black star) are indicated by white and yellow field lines, respectively; green (pink) field lines are rooted at the outer (inner) rim of the circular-shaped blue high-Q line in Panel (e). An animation of AIA 1600~{\AA} images is available in the online version of the journal.} 

\label{fig:2010-10-16mix1600_335_q_null}
\end{figure}

The ribbon morphology observed in the chromosphere (Figure~\ref{fig:2010-10-16mix1600_335_q_null}b) features a semi-circular ribbon (labeled R1), enclosing a short hook-shaped ribbon (labeled R2) in the east, and a remote extended ribbon (labeled R3) in the west. R1 and R2 are located in the center of the active region, running on either side of the circular-shaped polarity inversion line forming between the elongated positive fluxes and the surrounding negative fluxes (Figure~\ref{fig:2010-10-16mix1600_335_q_null}d). R3 is associated with the scattered positive fluxes in the facular region to the west. Based on a potential-field extrapolation using the $B_z$ component of the vector magnetogram, we calculate the map of the squashing factor $Q$ at the photosphere (Figure~\ref{fig:2010-10-16mix1600_335_q_null}e). A distinct feature in the $Q$-map is a circular-shaped high-Q line associated with negative polarity (blue) surrounding an elongated high-Q line associated with positive polarity (red). We find that the magnetic field lines (magenta) traced from the inner rim of the circular high-Q line are connected to the elongated high-Q line, while those (green) field lines traced from the outer rim of the circular high-Q line
are connected to a remote high-Q line in the west. Guided by these field lines, one can see that the ribbons R1, R2 and R3 are closely related to the circular, elongated, and remote high-Q lines, respectively (Figure~\ref{fig:2010-10-16mix1600_335_q_null}(b \& c)). Within the region bounded by the black box in Figure~\ref{fig:2010-10-16mix1600_335_q_null}d, we also identified a null point, whose spine (white) connects R2 and R3, while the field lines associated with the null's fan (yellow) are anchored in R1.

\begin{figure}[htbp]
\centering
\plotone{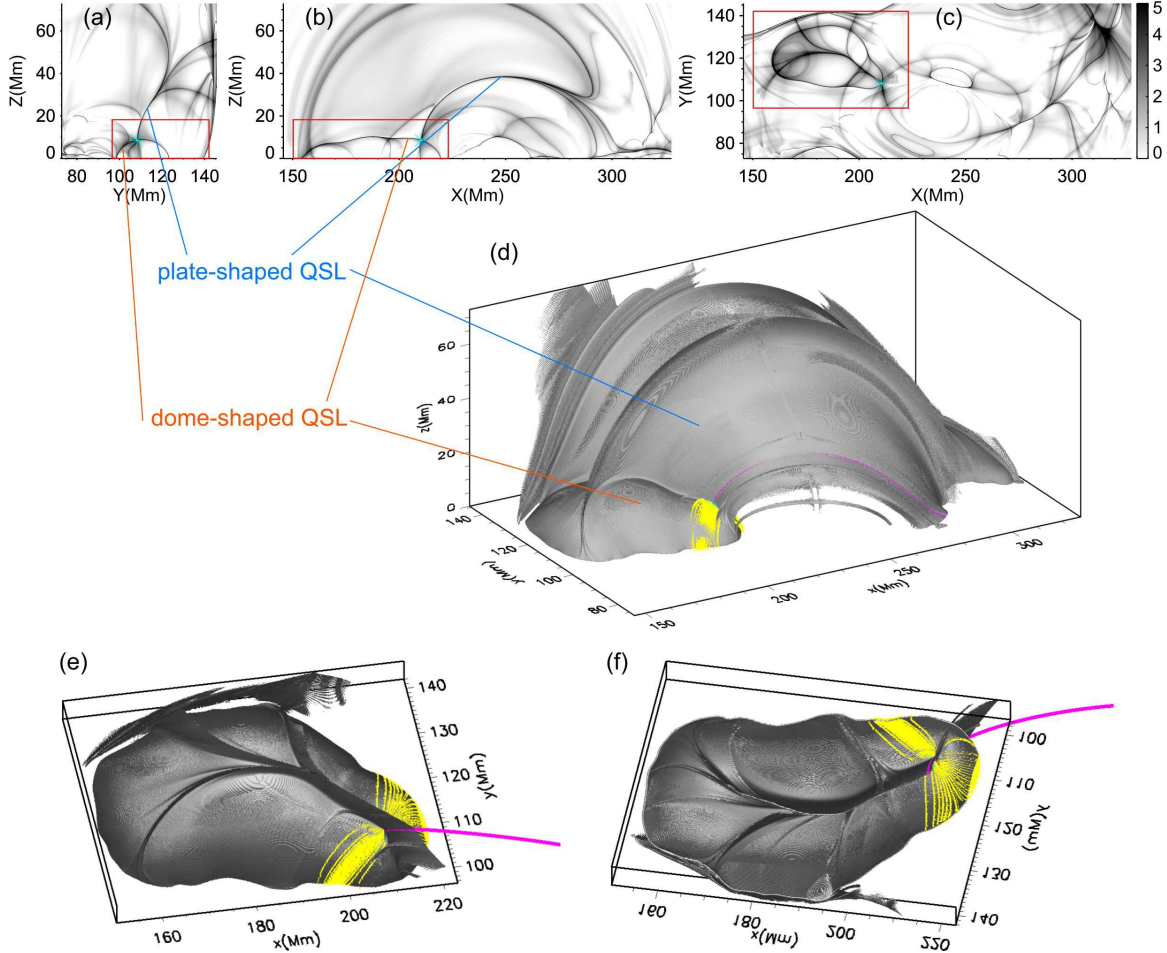}
\caption{Three-dimensional QSL related to the ELP flare on 2010 October 16. (a--c) Distribution of $\log_{10}Q$ at three cross-section(y-z, x-z, x-y) passing through a magnetic null point (cyan star), which is located at $(211.31, 108.22, 8.49)$~Mm. The origin of the coordinates is the same as in Figure~\ref{fig:2010-10-16mix1600_335_q_null}d. (d--f) Isosurface of $Q=10^4$ viewed from an oblique, top, and bottom perspective, respectively. Irrelevant structures that hinder the view have been manually removed. Panels (e-f) feature the dome-shaped QSL as bounded by the red boxes in (a-c). Purple (yellow) field lines show the spine (fan) originated from the null point. The CEA coordinates adopted here are the same as in Figure~\ref{fig:2010-10-16mix1600_335_q_null}(d--f). An animation the dome-plate QSL in 3D perspective is available in the online version of the Journal.
}
\label{fig:2010-10-16q3d}
\end{figure}

To understand the QSLs, which are three-dimensional structures, we calculate the squashing factor $Q$ within the 3D box region shown in Figure~\ref{fig:2010-10-16q3d}d. From the $Q$-maps in the cross sections ($y$-$z$, $x$-$z$, $x$-$y$) cutting through the null point (Figure~\ref{fig:2010-10-16q3d}(a-c)), one can see that the null point is located at the intersection of two QSLs. The circular high-Q line is the footprint of a dome-shaped QSL above the main-phase region, with three compartments in the dome (Figure~\ref{fig:2010-10-16q3d}(e \& f)). The elongated high-Q line is the footprint of a plate-shaped QSL intersecting the dome-shaped QSL and connecting with the remote high-Q line. The plate-shaped QSL embeds the spine field line. The inner and outer parts of this QSL were called the inner and outer spine-related QSL in \citet{Reid2012}. Hereafter, we refer to the composite QSL, featuring a plate intersecting a dome, as a ``dome-plate QSL''. The fan-spine structure of the null is a substructure of the dome-plate QSL: the spine is a single field line embedded within the plate-shaped QSL, and the fan is embedded within the western part of the dome-shaped QSL (Figure~\ref{fig:2010-10-16q3d}(d-f)). It is worth noting that the fan field lines extend only in part of the dome QSL and, correspondingly, their foot points do not cover the whole closed high-Q trace in the photosphere, mostly because the eigenvalues of the null point are significantly different from each other \citep{Parnell1996}. The footpoints of the fan field lines in Figure~\ref{fig:2010-10-16q3d} correspond well to the semi-circular ribbon R1 in Figure~\ref{fig:2010-10-16mix1600_335_q_null}(b). Comparing Figure~\ref{fig:2010-10-16q3d}d and Figure~\ref{fig:2010-10-16diff335}(a \& b), we found that the main-phase emission mainly originates from the dome-shaped QSL and the ELP emission mainly from the plate-shaped QSL.

\subsection{Characteristics of a two-ribbon ELP flare}\label{subsec:2r}

\begin{figure}[htbp]
	\centering
	\plotone{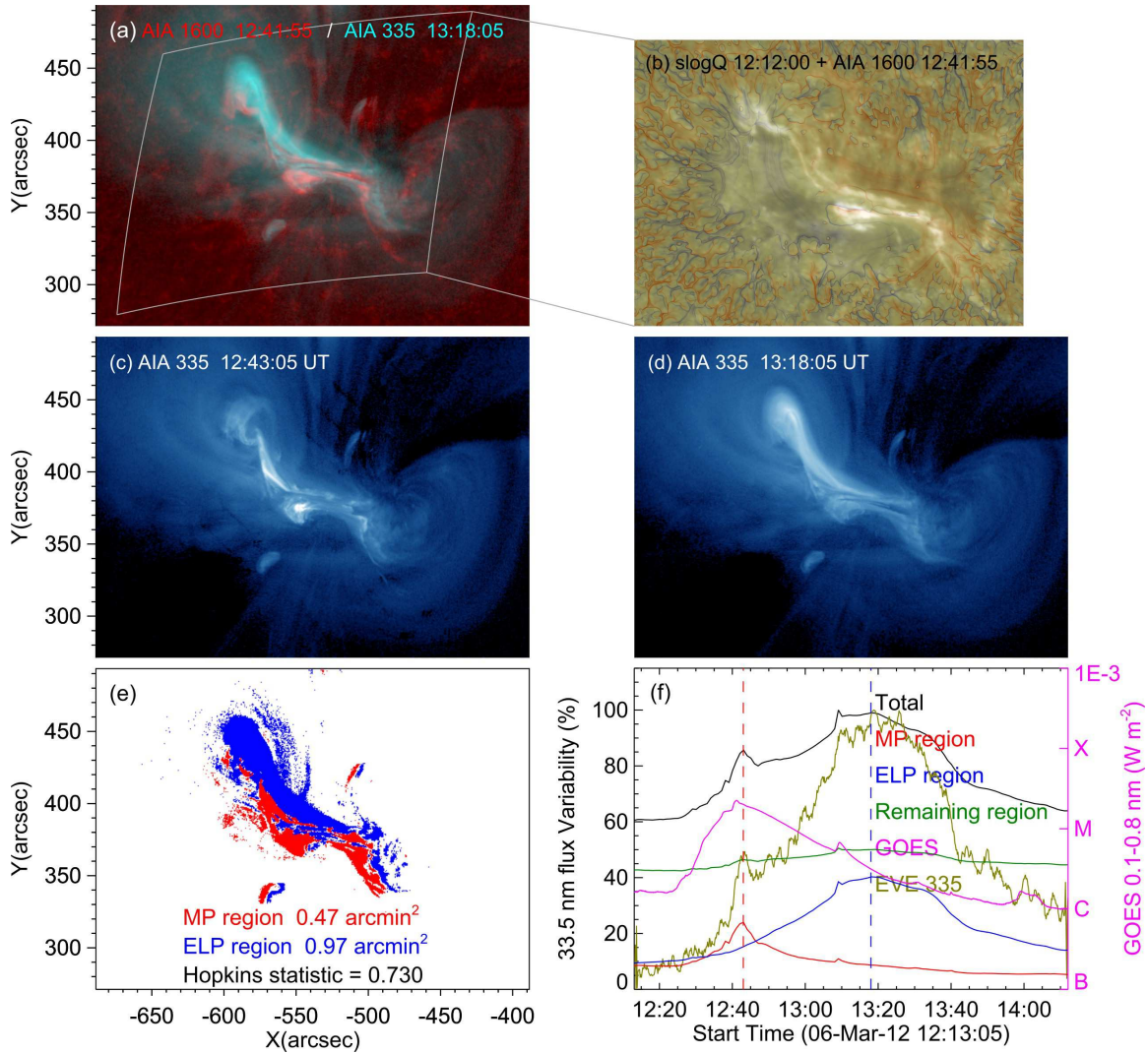}
	\caption{The two-ribbon ELP flare on 2013 March 6. (a) Composite image showing the 1600~{\AA} emission in red at 12:41:55 UT and 335~{\AA} in cyan at 13:18:05 UT. The curved parallelogram corresponds to the CEA-projected region shown in Panel (b). (b) Blend of the $Q$-map of potential field calculated from the photospheric $B_z$ at 12:12:00 UT and an AIA 1600~\AA\ image at 12:41:55~UT.	\deleted{The green field lines in (a) and (b) are traced from high-Q lines.} (c) AIA 335~{\AA} image at the main-phase peak. (d) AIA 335~{\AA} image at the ELP peak, with the same field of view as (c). (e) Main-phase and ELP regions are indicated by blue and red colors, respectively. (f) Light curves of \textsl{GOES} 1-8~{\AA} (scaled by the right $y$-axis), AIA 335~{\AA} data number integrated over the regions identified in Panel (e), and EVE 335~\AA\ (\ion{Fe}{16}) irradiance. The 335~{\AA} light curves are normalized in the same way as in Figure~\ref{fig:2010-10-16profile}. The vertical dashed red and blue lines mark the times of images in Panels (c) and (d), respectively.}
	\label{fig:two_example}			
\end{figure}

Surprisingly, many ELP flares are classic two-ribbon flares. Here we take the event occurring in NOAA AR 11429 on 2013 March 6, an extreme ELP event, as an example. The profile of the light curve in the AIA 335~\AA\ passband from the FOV covering the flaring region (Figure \ref{fig:two_example})  is very similar to the irradiance profile at EVE 335~\AA, hence we conclude that the EVE 335~\AA\ emission mainly comes from this active region. Moreover the AIA 335~\AA\ emission originates from essentially the same loop system during the main phase as during the ELP (Figure~\ref{fig:two_example}(c \& d)). Accordingly the Hopkins statistic (0.720) of this event is lower (Figure~\ref{fig:two_example}e) than that of the circular-ribbon flare on 2010 October 16. Further, unlike the circular-ribbon flare, the QSL footprints do not match the flare ribbons in AIA 1600~\AA\ (Figure~\ref{fig:two_example}b).

\subsection{Statistical Results}\label{subsec:stat}
Table~\ref{tab:StatisticResults} summarizes the statistical results for the sample of 55 ELP flares studied. What stands out is that the total number of two-ribbon flares is comparable to that of circular-ribbon flares. Even a larger proportion (52\%) of two-ribbon flares possess extreme ELP, i.e., the late-phase peak exceeding the main-phase peak, than that (37\%) of circular-ribbon flares. Obviously, the two-ribbon flares cannot be as well explained by magnetic skeletons calculated from pre-flare potential-field extrapolations as the circular-ribbon flares. It is also interesting that only about 1/3 of circular- and two-ribbon flares with ELP are associated with CMEs. For all three categories, 
the average Hopkins statistic is relatively high and on the same level, and the main-phase flaring region is only half as large as the ELP region. Both features indicate that the flaring loop system during the ELP is larger than that during the main phase, especially for circular-ribbon flares, which have the highest Hopkins statistic but lowest region size ratio among the three categories (Table~\ref{tab:StatisticResults}). A further unexpected result is that a magnetic null point and its associated fan-spine structure are only found in half of the circular-ribbon flares. The dome-plate QSL, however, can explain all but one of the circular-ribbon flares.

\begin{deluxetable}{c|ccc}

\tablecaption{
Statistical results of ELP flares 
\label{tab:StatisticResults}}
\tablehead{
\colhead{} &\colhead{Circular-ribbon} & \colhead{Two-ribbon} & \colhead{Complex-ribbon}
}
\startdata
  Number of events  &  19 & 23 & 13\\
  Eruptive events with CME(s) &7(37\%) & 8(35\%) & 7(54\%)\\
  Extreme ELP events &7(37\%) & 11(48\%) & 4(31\%)\\
  Flare ribbons matching QSL footprints &18(95\%) & 13(57\%) & 5(29\%)\\
  Fan-spine topology &10(53\%) & 2(9\%) & 1(8\%)\\ 
  $A_\mathrm{MP}/ A_\mathrm{ELP}$\tablenotemark{a}  & 0.397$\pm$0.251 & 0.456$\pm$0.186 &0.519$\pm$0.294\\  
  Hopkins statistic  & 0.780$\pm$0.064 & 0.720$\pm$0.075 &0.733$\pm$0.071\\
\enddata	
\tablenotetext{a}{Area ratio of main-phase over ELP region}
\end{deluxetable}

\begin{figure}[htbp]
	\plotone{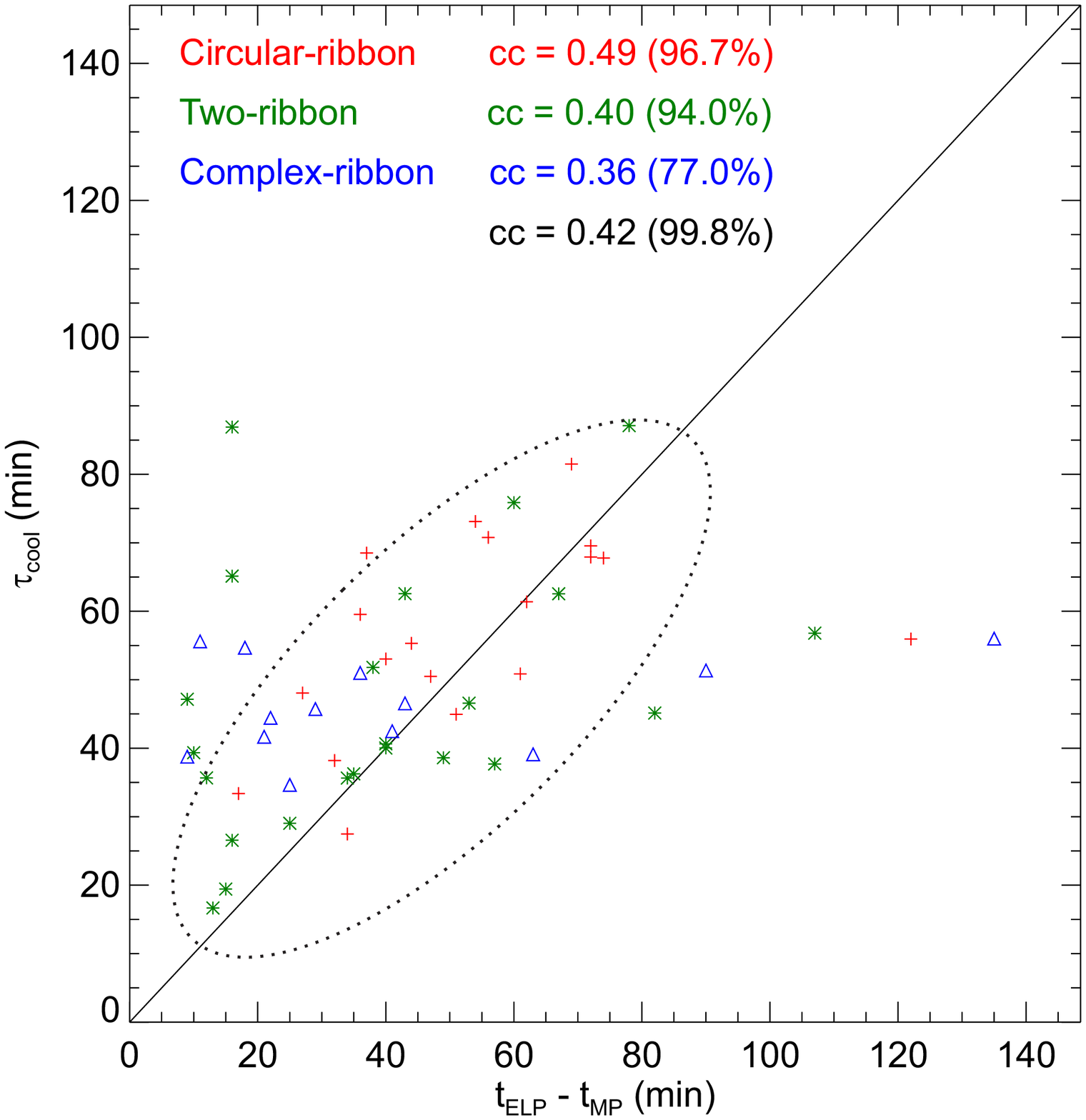}
	\caption{ELP in relation to plasma cooling. $t_\mathrm{ELP}-t_\mathrm{MP}$ is the time delay of the ELP peak relative to the main-phase peak in EVE 335~{\AA}. $\tau_\mathrm{cool}$ is the cooling timescale of flare plasma (Appendix~\ref{appendix:cooling_estimation}). Circular-ribbon, two-ribbon, and complex-ribbon flares are marked by red crosses, green asterisks, and blue triangles, respectively. The corresponding correlation coefficient (cc) is shown with the confidence level in the brackets. The cc in black is calculated for all of the events. For reference, the black solid line has the slope of unity. The dotted ellipse serves as a visual aid to mark the data points surrounding the reference line.}  \label{fig:coolings_statistic}
\end{figure}

\begin{figure}[htbp]
	\plotone{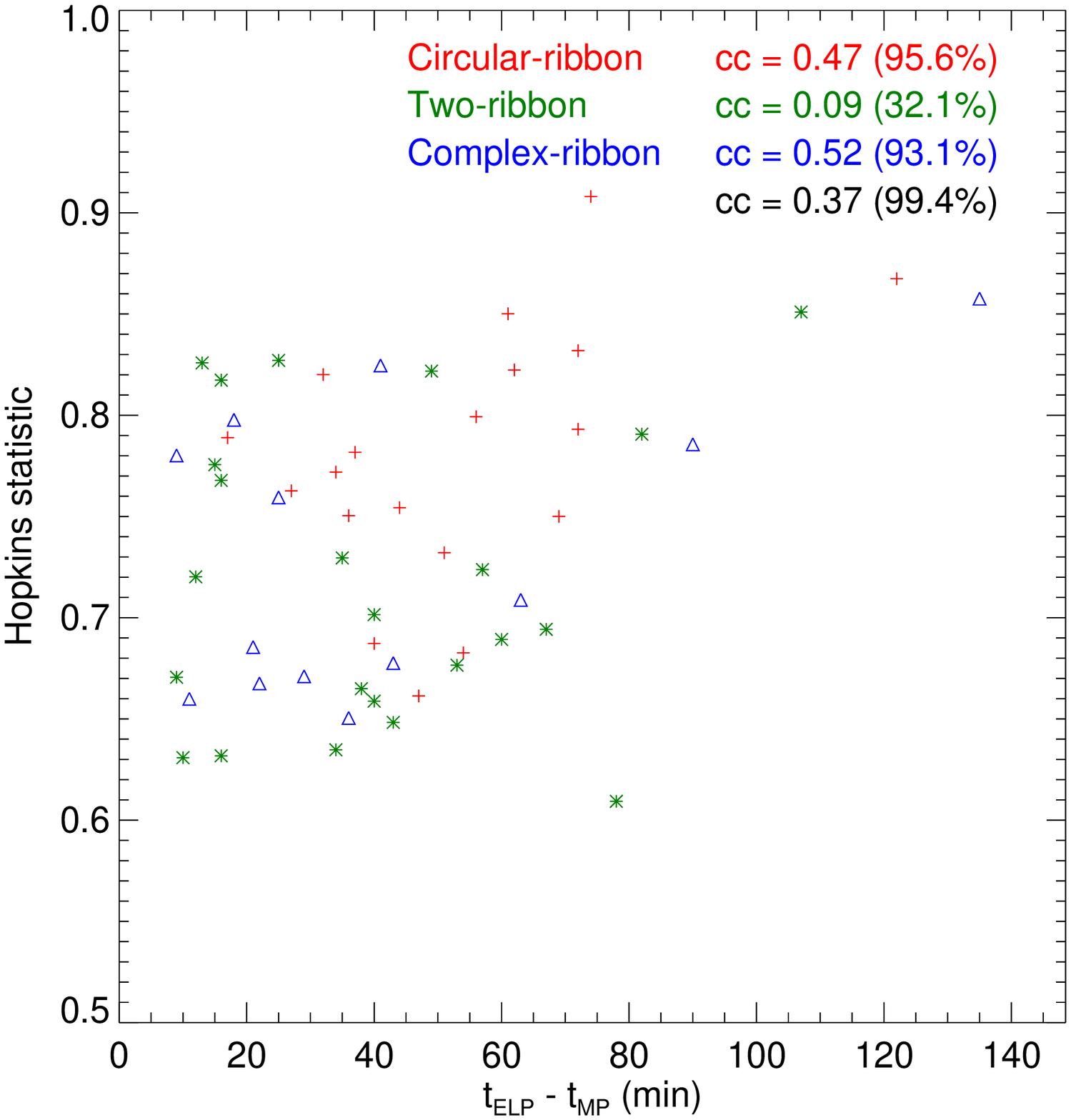}
\caption{Scatter plot of Hopkins statistic $H$ against $t_\mathrm{ELP}-t_\mathrm{MP}$. Circular-ribbon, two-ribbon, and complex-ribbon flares are marked by red crosses, green asterisks, and blue triangles, respectively. The corresponding cc is shown with the confidence level (using Student's $t$-distribution) in the brackets. The cc in black is calculated for all of the events.}  \label{fig:dt-Hopkins}
\end{figure}

\begin{figure}[htbp]
	\plotone{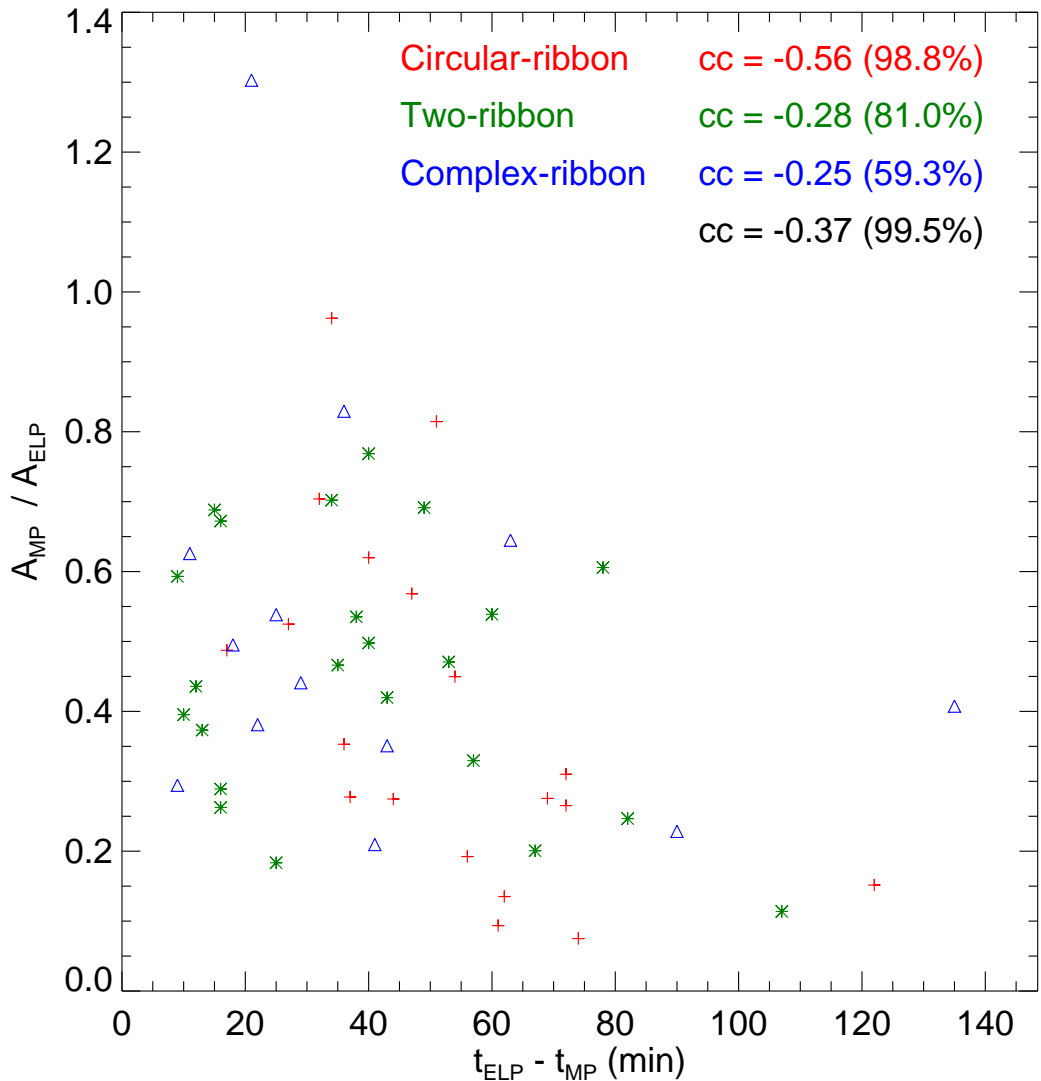}
	\caption{Scatter plot of the main-phase area over the ELP area $A_\mathrm{MP}/A_\mathrm{ELP}$ against $t_\mathrm{ELP}-t_\mathrm{MP}$. Circular-ribbon, two-ribbon, and complex-ribbon flares are marked by red crosses, green asterisks, and blue triangles, respectively. The corresponding cc is shown with the confidence level in the brackets. The cc in black is calculated for all of the events.}  \label{fig:dt-area_ratio}
\end{figure}

To understand the role of plasma cooling in ELP, we estimated cooling timescale $\tau_\mathrm{cool}$ of flare loops (Appendix~\ref{appendix:cooling_estimation}) to compare with the time delay of the ELP peak relative to the main-phase peak, $t_\mathrm{ELP}-t_\mathrm{MP}$, in the EVE 335~\AA\ irradiance profile (Figure~\ref{fig:coolings_statistic}). One can see that $\tau_\mathrm{cool}$ and $t_\mathrm{ELP}-t_\mathrm{MP}$ are correlated for both circular-ribbon and two-ribbon flares, at a two-sigma confidence level, and that except for a few outliers, most of which are two-ribbon and complex-ribbon flares, the majority of data points (marked by an ellipse) are located near the reference line marking $\tau_\mathrm{cool}=t_\mathrm{ELP}-t_\mathrm{MP}$. On the other hand, correlation coefficient is relatively low for complex-ribbon flares, with poor confidence level of the order of one-sigma. Moreover, we found that the Hopkins statistic $H$ is positively correlated with $t_\mathrm{ELP}-t_\mathrm{MP}$ (Figure~\ref{fig:dt-Hopkins}), but the area ratio of main-phase over ELP region $A_\mathrm{MP}/A_\mathrm{ELP}$ is negatively correlated with $t_\mathrm{ELP}-t_\mathrm{MP}$ (Figure~\ref{fig:dt-area_ratio}). Among the three flare categories, circular-ribbon flares show the highest correlation of $H$ and $A_\mathrm{MP}/A_\mathrm{ELP}$ against $t_\mathrm{ELP}-t_\mathrm{MP}$.

\begin{figure}[htbp]
	\plotone{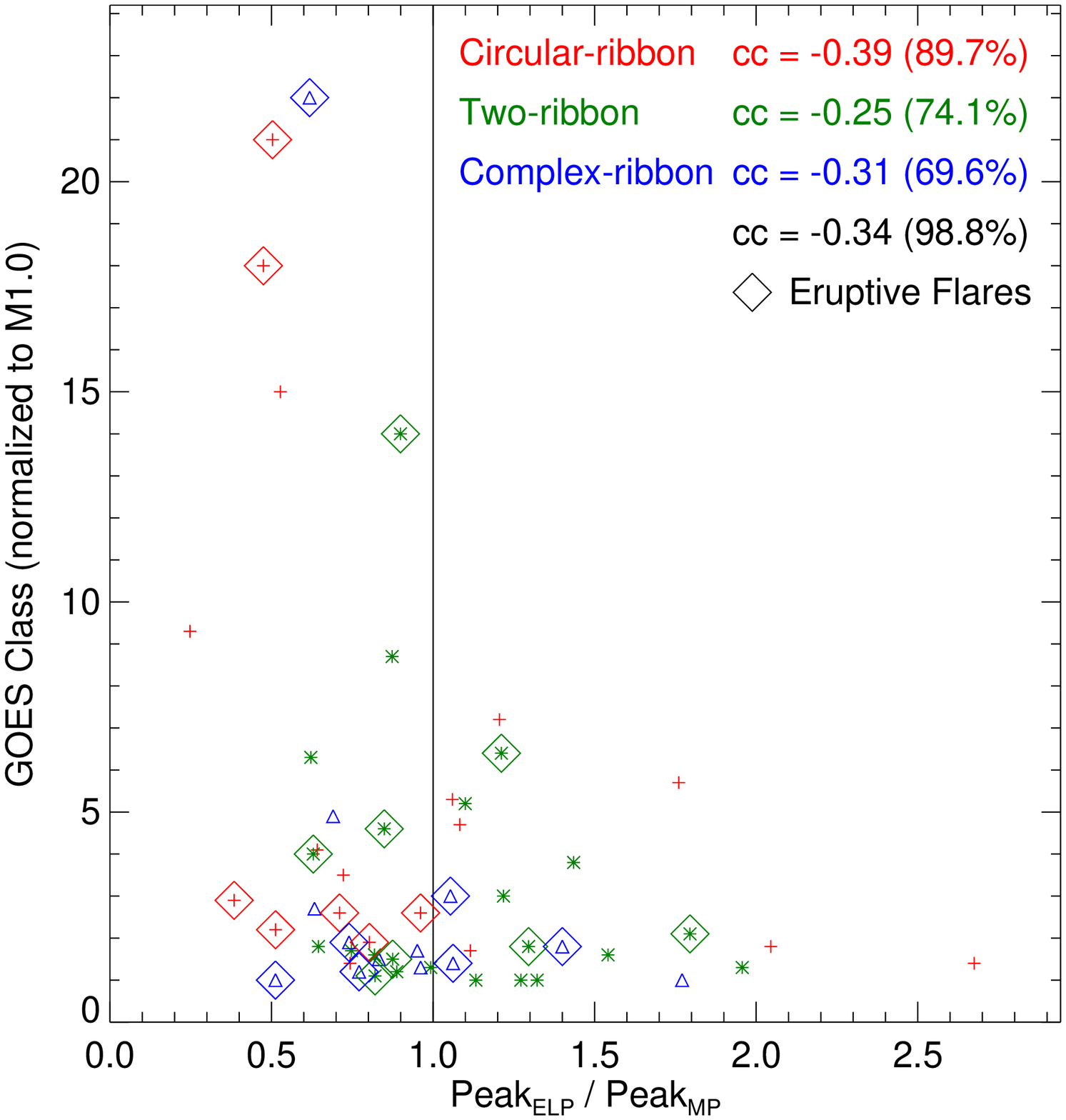}
	\caption{Scatter plot of GOES flare class against intensity ratio of the ELP peak over the main-phase peak at EVE 335~{\AA}. Circular-ribbon, two-ribbon, and complex-ribbon flares are marked by red crosses, green asterisks, and blue triangles, respectively. The corresponding cc is shown with the confidence level in the brackets. The cc in black is calculated for all of the events. The eruptive flares are marked by diamonds.}  \label{fig:peak_ratio}
\end{figure}

In addition, we found that the intensity ratio of the ELP peak over the main-phase peak shows a weak negative correlation with the flare class (Figure~\ref{fig:peak_ratio}). Only 6 of 22 eruptive but 16 of 33 confined flares are extreme ELP events, with the peak intensity ratio above unity. This suggests that extreme ELP events tend to be confined, which is consistent with a previous study using a much smaller sample of 12 ELP flares \citep{Wang2016}.

\section{Conclusions and Discussion}\label{sec:conclusion}
ELP events so far reported in the literature occurred predominantly in circular-ribbon flares. In our study spanning the period from May 2010 to May 2014, however, as many as 23 events of the total 55 M- and X-class flares with ELP are  classical two-ribbon flares, 19 events are circular-ribbon flares, and the flare morphology in the remaining 13 events is too complex to be simply categorized.  60\% of the ELP events are confined flares, about half of which have an extreme ELP.	 

In particular, an extreme ELP, i.e., the late-phase peak exceeding the main-phase peak, is found in 48\% of two-ribbon flares, 37\% of circular-ribbon flares, and 31\% of complex-ribbon flares (Table~\ref{tab:StatisticResults}), which suggests that additional heating during the ELP is more likely present in two-ribbon flares than in circular-ribbon flares. The origin of such heating is not yet clarified; however, \citet{Zhou2019} proposed a plausible model for events formed in quadrupolar source regions. On the other hand, extreme ELP events tend to be confined (Figure~\ref{fig:peak_ratio}), which may be related with heating associated with a trapped flux rope \citep{Wang2016}. Overall, we found that cooling is the dominant factor causing the delay of the ELP peak relative to the main-phase peak (Figure~\ref{fig:coolings_statistic}). This can be attributed to the fact that the loop system responsible for the ELP emission is generally larger than that responsible for the main-phase emission (Figures~\ref{fig:dt-Hopkins} \& \ref{fig:dt-area_ratio}). An important difference lies in the fact that, in the circular-ribbon flares, the two loop systems are well separated in space, while in the majority of the two-ribbon flares, the main phase and the ELP share essentially the same loop system that grows with time (Figure~\ref{fig:dt-Hopkins}). Thus, the ELP emission in two-ribbon flares is contributed by the ongoing magnetic reconnection taking place high in the corona \citep[see also][]{DaiYu2018,Zhou2019}.  

In addition, all of circular-ribbon flares are well accounted for by a ``dome-plate QSL''. The QSL embeds the fan-spine structure of a magnetic null point in about half of the events. The footprint of the dome-shaped QSL matches the circular-shaped ribbon, which is closely related to the main-phase emission. The plate-shaped QSL, consisting of long coronal loops connecting the remote ribbon with the inner ribbon, is mainly responsible for the ELP emission. A comparative investigation of the dome-plate QSL for source regions with and without a magnetic null point is beyond the scope of the present investigation and will be performed in future work. 

Two-ribbon flares, on the other hand, are poorly understood with structural skeletons derived from pre-flare potential-field extrapolations. This is not surprising because on one hand, magnetic topology is rapidly evolving during such eruptions, as manifested in the separating ribbons; and on the other hand, such flares are related to QSLs that are connected with the non-potential magnetic flux (a magnetic flux rope or a highly sheared arcade) in the source region. Our potential-field extrapolation does not reproduce such flux correctly, thus, may easily miss the relevant QSLs. In contrast, magnetic nulls in circular-ribbon flares are typically found in potential field high above the non-potential flux.

\appendix

\section{Flare list} \label{appendix:flarelist}
\startlongtable
\small
\begin{deluxetable*}{cccccccccc}
	\tablecaption{Flare list\label{tab:Flare list}}
	\tablecolumns{10}
	\tablehead{
		\colhead{Date} &
		\colhead{Onset} &
		\multicolumn{2}{c}{Location} & \colhead{Class}& \colhead{CME} &\colhead{Extreme ELP}&\colhead{QSL}&\colhead{Fan-Spine}&\colhead{Category} \\
		\cline{3-4}
		\colhead{} & \colhead{(UT)} &
		\colhead{AR} &\colhead{Position} & \colhead{} & \colhead{} 
	}
	\startdata  
2010-10-16 & 19:07 & 11112 & S20W26 & M2.9 & Y & N & Y & Y & Circular \\
2011-02-14 & 17:20 & 11158 & S20W04 & M2.2 & Y & N & N & N & Circular \\
2011-02-15 & 01:44 & 11158 & S20W10 & X2.2 & Y & N & Y & N & Complex \\
2011-03-09 & 23:13 & 11166 & N08W11 & X1.5 & N & N & Y & Y & Circular \\
2011-04-15 & 17:02 & 11190 & N14W19 & M1.3 & N & N & N & N & Two \\
2011-07-30 & 02:04 & 11261 & N14E35 & M9.3 & N & N & Y & N & Circular \\
2011-09-06 & 22:12 & 11283 & N13W18 & X2.1 & Y & N & Y & N & Circular \\
2011-09-07 & 22:32 & 11283 & N14W31 & X1.8 & Y & N & Y & N & Circular \\
2011-09-24 & 19:09 & 11302 & N12E42 & M3.0 & Y & Y & N & N & Complex \\
2011-09-26 & 14:37 & 11302 & N14E30 & M2.6 & Y & N & Y & Y & Circular \\
2011-09-28 & 13:24 & 11302 & N13E03 & M1.2 & Y & N & N & N & Complex \\
2011-11-05 & 20:31 & 11339 & N21E34 & M1.8 & N & Y & Y & Y & Circular \\
2011-11-06 & 06:14 & 11339 & N21E31 & M1.4 & N & N & Y & Y & Circular \\
2011-11-15 & 12:30 & 11346 & S19E32 & M1.9 & Y & N & Y & Y & Circular \\
2011-12-25 & 18:11 & 11387 & S22W26 & M4.0 & Y & N & Y & N & Two \\
2012-01-18 & 19:04 & 11401 & N17E32 & M1.7 & N & N & Y & N & Two \\
2012-01-23 & 03:38 & 11402 & N33W21 & M8.7 & N & N & N & N & Two \\
2012-03-06 & 12:23 & 11429 & N21E40 & M2.1 & Y & Y & N & N & Two \\
2012-05-09 & 14:02 & 11476 & N06E22 & M1.8 & N & N & Y & N & Two \\
2012-05-09 & 21:01 & 11476 & N12E26 & M4.1 & N & N & Y & Y & Circular \\
2012-05-10 & 04:11 & 11476 & N12E22 & M5.7 & N & Y & Y & Y & Circular \\
2012-05-10 & 20:20 & 11476 & N12E12 & M1.7 & N & Y & Y & N & Circular \\
2012-06-14 & 12:52 & 11504 & S19E06 & M1.9 & Y & N & N & N & Complex \\
2012-07-04 & 09:47 & 11515 & S17W18 & M5.3 & N & Y & Y & Y & Circular \\
2012-07-04 & 22:03 & 11515 & S16W28 & M4.6 & Y & N & Y & N & Two \\
2012-07-05 & 03:25 & 11515 & S18W29 & M4.7 & N & Y & Y & Y & Circular \\
2012-07-06 & 08:07 & 11514 & S17W40 & M1.5 & N & N & Y & N & Complex \\
2012-07-06 & 10:24 & 11515 & S17W42 & M1.8 & Y & Y & Y & N & Complex \\
2012-07-12 & 15:37 & 11520 & S13W03 & X1.4 & Y & N & Y & N & Two \\
2012-07-14 & 04:51 & 11520 & S22W36 & M1.0 & N & Y & N & N & Complex \\
2012-09-08 & 17:35 & 11564 & S14W40 & M1.4 & N & Y & Y & N & Circular \\
2012-11-20 & 19:21 & 11618 & N07E15 & M1.6 & N & N & N & N & Two \\
2012-11-27 & 21:05 & 11620 & S14W41 & M1.0 & N & Y & Y & N & Two \\
2013-01-11 & 14:51 & 11654 & N06E39 & M1.0 & Y & N & Y & N & Complex \\
2013-08-12 & 10:21 & 11817 & S21E17 & M1.5 & Y & N & N & N & Two \\
2013-10-15 & 08:26 & 11865 & S21W14 & M1.8 & Y & Y & N & N & Two \\
2013-10-15 & 23:31 & 11865 & S21W22 & M1.3 & N & N & N & N & Complex \\
2013-10-22 & 00:14 & 11875 & N06E16 & M1.0 & N & Y & Y & Y & Two \\
2013-10-23 & 20:41 & 11875 & N05W06 & M2.7 & N & N & Y & N & Complex \\
2013-10-24 & 10:30 & 11875 & N06W12 & M3.5 & N & N & Y & N & Circular \\
2013-11-01 & 19:46 & 11884 & S12E01 & M6.3 & N & N & Y & N & Two \\
2013-11-03 & 05:16 & 11884 & S12W17 & M4.9 & N & N & N & N & Complex \\
2013-11-16 & 07:45 & 11900 & S20W30 & M1.6 & N & Y & N & N & Two \\
2013-12-31 & 21:45 & 11936 & S15W36 & M6.4 & Y & Y & Y & N & Two \\
2014-01-07 & 10:07 & 11944 & S13E13 & M7.2 & N & Y & Y & N & Circular \\
2014-01-31 & 15:32 & 11968 & N07E34 & M1.1 & Y & N & Y & N & Two \\
2014-02-01 & 01:19 & 11967 & S10E26 & M1.0 & N & Y & Y & N & Two \\
2014-02-01 & 07:14 & 11967 & S10E22 & M3.0 & N & Y & Y & N & Two \\
2014-02-02 & 06:24 & 11968 & N12E18 & M2.6 & Y & N & Y & N & Circular \\
2014-02-04 & 01:16 & 11967 & S09W13 & M3.8 & N & Y & Y & N & Two \\
2014-02-04 & 03:57 & 11967 & S14W07 & M5.2 & N & Y & N & N & Two \\
2014-02-05 & 16:11 & 11967 & S10W36 & M1.3 & N & Y & N & N & Two \\
2014-02-13 & 15:45 & 11974 & S11W16 & M1.4 & Y & Y & N & Y & Complex \\
2014-03-03 & 15:54 & 11989 & N07W36 & M1.2 & N & N & N & Y & Two \\
2014-03-10 & 15:21 & 12002 & S17E41 & M1.7 & N & N & N & N & Complex \\
	\enddata
	\tablecomments{
		``Y'' for Yes, ``N'' for No. The column ``QSL'' indicates whether the flare ribbons match the QSL footprints. The column ``Fan-Spine'' indicates whether there exists a relevant coronal null with the fan-spine topology.		
	}
\end{deluxetable*}

\section{Hopkins Statistic}\label{appendix:hopkins}

The Hopkins statistic \citep{Hopkins1954,Banerjee2004} measures the cluster tendency of a data set. A set of highly clustered points give a Hopkins statistic close to 1, a set of randomly distributed points tend to produce a Hopkins statistic around 0.5, and a set of uniformly distributed points yield a Hopkins statistic close to 0. In this work, we randomly sample a set of $n$ points $P_1$ from the main-phase region, and the same number of points $P_2$ from the ELP region ($n=100$ in our case). The Hopkins statistic $H$ is calculated as follows, 
\begin{equation}
H=\frac{1}{2}\left(\frac{\Sigma^n_{i=1}d_{21,i}}{\Sigma^n_{i=1}d_{21,i}+{\Sigma}^n_{i=1}d_{11,i}}+ \frac{\Sigma^n_{i=1}d_{12,i}}
{\Sigma^n_{i=1}d_{12,i}+{\Sigma}^n_{i=1}d_{22,i}}\right),
\end{equation}
where $d_{\alpha\beta,i}$ indicates the linear distance of a point $i\in$ $P_\alpha$ from its nearest neighbor in $P_\beta$, with $\alpha,\beta=1,2$. The original definition of Hopkins statistic considers only the first term inside the brackets of the above equation, but the two terms generally differ. Hence we take their average as the final assessment.

\section{Estimation of cooling timescale }\label{appendix:cooling_estimation}

Following \citet{Cargill1995}, we estimated the cooling timescale of flare plasma as follows,
\begin{eqnarray}
\tau_\mathrm{cool}=2.35 \times 10^{-2}\, L^{5/6}\,T_e^{-1/6}\,n_e^{-1/6},
\label{eqn:cooling_orig}
\end{eqnarray}
which assumes that initially conductive cooling dominates, but later on radiative cooling takes over until cooling down to $\sim\,$1~MK. The initial electron temperature $T_e$ is estimated by the ratio of the two \textsl{GOES} SXR passbands, 0.5--4~{\AA} and 1--8~{\AA}, at the flare peak. The electron density $n_e$ is estimated from the \textsl{GOES} emission measure, i.e., $EM=\int n_e^2 dV \thickapprox n_e^2 V$. Instead of measuring the loop half length, we took on $L$ as the spatial scale of the ELP loop system and $V\sim L^3$. We estimated $L$ by the square root of the flaring area obtained in the same way as the ELP region in Figure~\ref{fig:2010-10-16diff335}c. 

To compare $\tau_\mathrm{cool}$ with the time delay of the ELP peak relative to the main-phase peak, i.e., $t_\mathrm{ELP}-t_\mathrm{MP}$ in Figure~\ref{fig:coolings_statistic}, we modified $\tau_\mathrm{cool}$ to take into account the further cooling from the ELP-peak temperature down to about 1~MK. This time period must be dominated by radiative cooling ${\tau}_\mathrm{rad}=3\, k_Bn_eT_\mathrm{e} / P_\mathrm{rad}$, where the optically thin radiative loss function $P_\mathrm{rad}$ between 1--10~MK is approximated by $1.2\times 10^{-19} n_e^2T_e^{-1/2}$ \citep{Cargill1995}. Thus, $\tau_\mathrm{cool}$ in Figure~\ref{fig:coolings_statistic} has been further subtracted by $3.45\times 10^3T_e^{3/2}/n_e$, where $T_e=2.5$~MK is the formation temperature of EVE 335~{\AA}.

\acknowledgements  
RL and BK acknowledge the NSFC-DFG collaborative grant NSFC 41761134088 and KL 817/8-1. This work was also supported by NSFC 41774150, 11925302, and 41421063, CAS Key Research Program KZZD-EW-01-4, and the fundamental research funds for the central universities. JC acknowledges support by the China Scholarship Council (No. 201706340140).

 \end{document}